\documentclass[thmsb]{article}
\usepackage{graphicx}
\usepackage{amsmath}
\usepackage{hyperref}
\input amssym.def
\input amssym

\newtheorem{theorem}{Theorem}

\newtheorem{lemma}{Lemma}

\newtheorem{proposition}{Proposition}
\newenvironment{proof}[1][Proof]{\textbf{#1.} }{\ \rule{0.5em}{0.5em}}

\begin{document}

\title{Unsharp Localization and Causality\\
in Relativistic Quantum Theory}
\author{Paul Busch \\
{\small Former address: Department of Mathematics, The University of Hull, Hull, UK}\\
{\small Current address: Department of Mathematics, University of York, York, UK}\\
{\small Electronic mail: {\tt paul.busch@york.ac.uk}}\\
\hspace{1cm}\hfill\\
{\small Published in: J. Phys. A: Math. Gen. 32 (1999) 6535-6546}\\
{\small DOI: \href{http:dx.doi.org/10.1088/0305-4470/32/37/305}{10.1088/0305-4470/32/37/305}}
}
\date{}
\maketitle
\begin{abstract}
\noindent The conflict between relativistic causality and localizability is
analyzed in the light of the existence of unsharp localization observables.
A theorem due to S.~ Schlieder is generalized, showing that the assumption
of local commutativity implies the localization observable in question to be unsharp in
a strong sense. Furthermore, a recent generalization of a theorem of
L\"uders is applied to demonstrate that local commutativity is a necessary consequence
of Einstein causality even in the case of unsharp observables if they
admit local measurements. These findings raise the question whether localization
observables can be measured by means of local operations.
\end{abstract}

\section{Introduction}

The concept of localization raises intriguing problems in relativistic
quantum theory. On one hand, the idea of localizability (of particles,
centres of charge or energy distributions, etc.) has always had an
unquestionable heuristic and interpretational value. On the other hand, any
attempt at a formalization of localization as an observable seems to face a
fundamental conflict with the requirement of causality on which
(together with some other postulates) relativistic quantum theories are
built. This conflict is epitomised in theorems due to S.\ Schlieder \cite
{S71} and G.\ Hegerfeldt \cite{H98}.

Further problems arise in the context of the relativistic quantum
mechanics of (free) particles. For example, it has been noted that a
conserved probability current with positive probability density does not
exist in all cases (e.g., \cite{Schw61,Gr94}). Furthermore, no sharp
localization observable exists for particles with zero mass and 
spin at least one \cite{Wigh62}. These problems are overcome in an approach that
describes (spatial) localization in terms of marginal observables of 
covariant phase space observables (e.g., \cite{Ali85,Ali98,BrS96}). Such
observables are unsharp observables represented as noncommutative
positive operator measures (\textsc{pom}'s), thus accounting for
the noncommutativity of position and momentum and the uncertainty
relation. This success raises the question whether localizability and
causality can be reconciled for unsharp (spatial) localization
observables. A pragmatic answer (``FAPP") has been given in the phase space
approach in \cite{GP84}, indicating that the possibility for causality
violating behavior is spurious. 

In the present paper Schlieder's theorem is reconsidered in order to decide
whether its statement also holds in the case of unsharp observables. 
It turns out that localization observables will necessarily be unsharp in
a strong sense if they are to satisfy the local commutativity condition
(Section 2). The question of whether the local commutativity of 
\textsl{unsharp} localization observables is in turn a necessary consequence
of Einstein causality is addressed in Section 3. General conclusions and
problems for future investigation are summarized in Section 4. For the
readers' convenience, basic concepts relating to unsharp observables
represented as \textsc{pom}'s are briefly reviewed in  Appendix 1.

\section{Unsharp Localization vs. Local Commutativity}

Schlieder's theorem \cite{S71} is based on the following structures 
commonly accepted as fundamental for any relativistic quantum theory.

\begin{description}
\item[(a)]  a complex Hilbert space $\mathcal{H}$, the rays of which
represent pure states of the system;

\item[(b)]  a strongly continuous unitary representation $a\mapsto
U(a)$ in $\mathcal{H}$ of the translation group of Minkowski space $M$;

\item[(c)]  for any future directed, timelike unit vector $a$, the generator 
$H(a)$ (Hamiltonian) is bounded below (spectrum condition).
\end{description}

\noindent Within the structure $\left( \mathcal{H},a\mapsto
U(a)\right)$, the following conditions are taken to characterize a
(spatial) localization observable within a given inertial frame.

\begin{description}
\item[(0)]  \textsl{Localization Event Structure:} Fix a foliation of
Minkowski space $M$ by means of a family $\mathcal{S}$ of parallel spacelike
hyperplanes $S$; each $S$ is required to be equipped with a family 
$\mathcal{F}(S)$ of subsets, called \textsl{spatial sets}, including a (covering)
family of bounded subsets, and such that $\mathcal{F}(S+a)$ consists of the
translates by $a$ of the sets from $\mathcal{F}(S)$; and a map $\Delta
\mapsto E_{\Delta }$ from $\mathcal{F}(S)$ to effects of $\mathcal{H}$
for each $S$.

\item[(1)]  \textsl{Translation Covariance:} For all $a\in M$, 
\[
U(a)E_{\Delta }U(a)^{\ast }=E_{\Delta +a}.
\]

\item[(2)]  \textsl{Localizability:} For each
$S\in{\mathcal{S}}$ and $\Delta _{1},\Delta
_{2}\in \mathcal{F}(S)$, 
\[
\mathrm{if\ }\Delta _{1}\cap \Delta _{2}=\emptyset\quad
\mathrm{then}\quad 
E_{\Delta _{1}}E_{\Delta _{2}}=E_{\Delta _{2}}E_{\Delta _{1}}=0.
\]

\item[(3)]  \textsl{Local Commutativity:} For $S_1,S_2\in\mathcal{S}$,  
$\Delta _{1}\in \mathcal{F}(S_{1}),\Delta _{2}\in \mathcal{F}(S_{2})$, 
\[
\mathrm{if\ }\Delta _{1},\Delta _{2}\ \mathrm{spacelike \ separated\quad then\quad}
E_{\Delta _{1}}E_{\Delta _{2}}=E_{\Delta _{2}}E_{\Delta _{1}}.
\]
\end{description}

\noindent The theorem then reads:

\begin{theorem}
\label{t1}If the structure $\left( \mathcal{H},a\mapsto U(a),\Delta \mapsto
E_{\Delta }\right) $, with the $E_{\Delta}$ being projections $P_{\Delta}$,
satisfies conditions (1)-(3), then $P_{\Delta }=0$ for
all bounded spatial sets $\Delta $.
\end{theorem}

\noindent The present formulation of the theorem for projection valued
maps $\Delta\mapsto P_{\Delta}$ is due to D.~Malament \cite{Mal96}.
In Schlieder's original version, condition (3)
was replaced with the somewhat stronger requirement (referred to as a
consequence of causality and similar to Hegerfeldt's \cite{H98}
characterization of causality) that the product 
$P_{\Delta_{1}}P_{\Delta _{2}}=0$ for spacelike separated pairs of sets; then it
is pointed out that the localization projections $P_{\Delta }$ cannot
belong to any local algebra. Examples of such `nonlocal' localization
observables are given by the Newton-Wigner position operators
\cite{NW49} or the corresponding localization spectral measures
constructed by Wightman  \cite{Wigh62}. It should be noted that these
sharp localization observables do satisfy the covariance condition (1)
\cite{CM82}.

Theorem 1 has been interpreted as implying the impossibility of a
physically acceptable notion of localizability of physical systems in
relativistic quantum theories. Inasmuch as a localization observable is a
defining feature of a particle, this conclusion would entail the
impossibility of a relativistic quantum mechanics of particles. 
However, before subscribing to such far-reaching conclusions, it is worthwhile to
reanalyze the assumptions in some detail in view of the possibility that
localization might be an intrinsically unsharp observable.

In what follows we will not question the postulates (a), (b) and (c),
nor the property (1) of translation covariance. We will also tentatively
adopt the view that the operational definition of spatial sets and the
continuum of possible spacetime translations can be realized with
arbitrary accuracy with classical physical, macroscopic means. The
implications of the fact that measuring devices such as detectors are
composed of quantum constituents will only be explored in the discussion
at the end.

The formulation of the localization event structure is usually somewhat
sharpened by requiring the maps $\Delta\mapsto E_{\Delta}$ to be 
\textsc{pom}'s defined on spatial Borel sets. As we shall see, this would 
simplify some of the arguments below. Physically, it would reflect the
assumption that localization is operationally meaningful for arbitrarily
small sets, which could again be challenged on the grounds that
detectors are composed of quantum systems.

First we investigate the question whether or not the statement of
Theorem 1 extends to unsharp localization observables as well. 
To this end we will allow the structure 
$\left( \mathcal{H},a\mapsto U(a),\Delta \mapsto E_{\Delta }\right)$ 
to be based on effects $E_{\Delta }$ which are not necessarily projections.

\begin{theorem}
\label{t2}If the structure $\left( \mathcal{H},a\mapsto U(a),\Delta \mapsto
E_{\Delta }\right) $ satisfies conditions (1)-(3), then $E_{\Delta }=0$ for
all (bounded) spatial sets $\Delta $.
\end{theorem}

\noindent We will sketch a proof of this statement as it is convenient to
display the technicalities involved -- they are exactly the same as in the
original case. As in that case, the proof rests on the following substantial
result due to Borchers \cite{Bor67}, presented here in terms of effects
rather than projections only.

\begin{lemma}
\label{l1}Let $V(t)$ be a strongly continuous one-parameter group of unitary
operators on a Hilbert space whose generator $H$ has a spectrum bounded from
below. Let $E_{1},E_{2}$ be two effects such that \newline
(i) $E_{1}E_{2}=0$, and\newline
(ii) there is $\varepsilon >0$ such that for all $t$ with $|t|<\varepsilon$,
$[E_{1},V(t)E_{2}V(t)^{\ast }]=0$.\newline
Then $E_{1}V(t)E_{2}V(t)^{\ast }=0$ for all $t\in \Bbb{R}$.
\end{lemma}

\noindent
\begin{proof}
We only show how the statement can be reduced to the known one for
projections. Let $P_{1},P_{2}$ be the projections onto the ranges of 
$E_{1},E_{2}$, respectively. Then $P_{1}^{(0)}=I-P_{1}$, $P_{2}^{(0)}=I-P_{2}$
are the projections onto the kernels. Now, observe that $E_{1}E_{2}=0
\Leftrightarrow P_{1}P_{2}=0$. [In fact: $E_{1}E_{2}\varphi =0\forall
\varphi \Leftrightarrow \mathrm{ran}(E_{2})\subseteq \ker
(E_{1})\Leftrightarrow P_{2}\leq P_{1}^{(0)}\Leftrightarrow
P_{2}(I-P_{1})=P_{2}\Leftrightarrow P_{2}P_{1}=0$.] Next observe that 
$V(t)P_{2}V(t)^{\ast }$ is the projection onto the range of 
$V(t)E_{2}V(t)^{\ast }$. Hence, (ii) implies $[P_{1},V(t)P_{2}V(t)^{\ast }]=0$
for all $t$ with $|t|<\varepsilon $. So we have (i) and (ii) for 
$P_{1},P_{2} $, and therefore $P_{1}V(t)P_{2}V(t)^{\ast }=0$ for all $t$. But
this is equivalent to $E_{1}V(t)E_{2}V(t)^{\ast }=0$ for all $t$.
\end{proof}

\noindent It is straightforward to check that Malament's line of argument
goes through for effects $E_{\Delta }$ in place of projections $P_{\Delta }$,
yielding $E_{\Delta }=0$ for all bounded $\Delta $. We will not carry this
out here but later in a somewhat less trivial context. Here we note the
following. In view of the probabilistic interpretation of the operators 
$E_{\Delta }$, if $\Delta _{1},\Delta _{2}\in \mathcal{F}(S)$ are two
disjoint subsets, then the sum of expectations of $E_{\Delta _{1}}$ and 
$E_{\Delta _{2}}$ should represent the probability of localizing the system
in $\Delta _{1}\cup \Delta _{2}$, and the operator representing these
probabilities is $E_{\Delta _{1}\cup \Delta _{2}}$. Thus one is led to
stipulate the additivity of the map $\Delta \mapsto E_{\Delta }$, that is, 
$E_{\Delta _{1}}+E_{\Delta _{2}}=E_{\Delta _{1}\cup \Delta _{2}}$ for
disjoint $\Delta _{1},\Delta _{2}\in \mathcal{F}(S)$. In addition, one might
consider the possibility that localization \textsl{somewhere} in a given
hyperplane would occur with certainty, that is, $E_{S}=I$. We note that for
these probabilistic requirements to be implemented, it is necessary that the
family $\mathcal{F}(S)$ is an algebra. [One usually assumes it to be a a 
$\sigma $-algebra, such as, for instance, the Borel algebra of $S$.] But then
for a normalized \textsc{pom}, the localization condition (2) is
seen to be equivalent to the effects $E_{\Delta }$ being projections. In
fact (2) implies $E_{\Delta }(I-E_{\Delta })=0$, that is, $E_{\Delta
}^{2}=E_{\Delta }$. The converse implication is trivial. So if localization
observables were adequately represented as normalized \textsc{pom}'s,
the contents of Theorem \ref{t2} would immediately reduce to that of the
original theorem. But one should keep in mind that the normalization
condition does not apply to all \textsc{pom}'s representing physical
observables.

Our main criticism of Theorem \ref{t2} \ aims at another aspect: the
formalization of the localization condition in terms of the algebraic
condition $E_{\Delta _{1}}E_{\Delta _{2}}=0$ for disjoint spatial sets. What
one actually tries to express with this condition is the following: ``If the
system is in $\Delta _{1}$, it certainly is not in $\Delta _{2}$ whenever
these sets are disjoint.'' Thus, (2) should be replaced with

\begin{description}
\item[(2')]  For all states $\varphi \in \mathcal{H},\Vert \varphi \Vert =1$,
$\Delta _{1},\Delta _{2}\in \mathcal{F}(S)$, 
\[
\mathrm{if\ }\Delta_{1}\cap \Delta _{2}=\emptyset \quad\mathrm{then}\quad
\langle \varphi |E_{\Delta _{1}}\varphi \rangle =1\Longrightarrow \langle
\varphi |E_{\Delta _{2}}\varphi \rangle =0. 
\]
\end{description}

\noindent This is equivalent to $P_{\Delta _{1}}^{(1)}\leq P_{\Delta
_{2}}^{(0)}$, where $P_{\Delta }^{(1)},$ $P_{\Delta }^{(0)}$ denote the
spectral projections of $E_{\Delta }$ associated with the eigenvalues 1 and
0, respectively. It is obvious that this condition is equivalent to (2) in
the case of projections. For effects, (2') only implies $P_{\Delta
_{1}}^{(1)}$ $P_{\Delta _{2}}^{(1)}=0$ while in general $E_{\Delta
_{1}}E_{\Delta _{2}}\neq 0$. Note that (2') can be obtained as a consequence
of the assumption that $\Delta \mapsto E_{\Delta }$ is a (not necessarily
normalized) \textsc{pom}: if,$\ $for disjoint $\Delta _{1},\Delta
_{2}$, $E_{\Delta _{1}}+E_{\Delta _{2}}=E_{\Delta _{1}\cup \Delta _{2}}(\leq
I)$ and $\langle \varphi |E_{\Delta _{1}}\varphi \rangle =1$, then $\langle
\varphi |E_{\Delta _{2}}\varphi \rangle =0$.

\begin{theorem}
\label{t3}If the structure $\left( \mathcal{H},a\mapsto U(a),\Delta \mapsto
E_{\Delta }\right) $ satisfies conditions (1), (2') and (3), then $P_{\Delta
}^{(1)}=0$ for all (bounded) spatial sets $\Delta $.
\end{theorem}

\noindent In  Appendix 2 we show how to adjust the proof of Theorem
\ref{t1} to obtain this result. We interpret this result as follows: if
there exists a localization observable satisfying the conditions of
Theorem \ref{t3} then this observable is necessarily \textsl{strongly
unsharp} in the sense that its effects $E_{\Delta}$ do not have
eigenvalue 1. This leaves us with the question whether among the
strongly unsharp, covariant localization observables there exist any
that satisfy local commutativity. To my knowledge the answer is not
known. An indication to the negative is provided by a recent theorem
stating that \textsl{spacetime} localization observables cannot belong
to any quasilocal algebra \cite{Gia98}.

It is known that among the relativistic phase space observables there are
strongly unsharp observables, whose spatial marginals would
thus in general be strongly unsharp covariant localization observables.
In the remaining part of the paper we address the question
of the relevance of the local commutativity condition (3) in the case of
unsharp localization observables: is it a necessary consequence of the
requirement of Einstein causality? We shall find a partial answer and along
the way obtain an indication that phase space localization observables do
violate local commutativity.

\section{Local Commutativity and Causality for Unsharp Localization Observables}

The term ``Einstein causality" refers to the intuitive idea of physical
processes propagating with at most the velocity of light. Following
Schlieder \cite{S71} and Heger\-feldt \cite{H98}, we will distinguish between
\textsl{weak} and \textsl{strong} (Einstein) causality: the former
refers to subluminal propagation of changes of expectation values,
whereas the latter describes subluminal propagation of individual,
definite properties. Alternative terms in use are \textsl{macro-} and
\textsl{microcausality}, respectively.
 
For sharp observables, the local commutativity condition is known to be
equivalent to weak Einstein causality \cite{S68}. This result is a consequence of a
famous theorem due to L\"uders \cite{Lud51} which states the following:
two (discrete) observables represented by self-adjoint operators $A,B$
commute if and only if for any state, the statistics of a measurement of
$B$ is not affected by a nonselective \textsl{L\"{u}ders measurement }of $A$
(that is, a measurement without
reading). This theorem can be extended to observables whose spectra are
not discrete: the commututativity $A,B$ is equivalent to the statistical nondisturbance
condition being stipulated for L\"uders measurements on all discrete coarse-grainings of the
observable $A$. 
Now let $A,B$ be observables that can be measured in  two
spacelike separated regions of spacetime, respectively. Then weak
Einstein causality requires that acts of  measurements in
one region should not have statistically significant effects in another
region at a spacelike separation. This is captured by the
nondisturbance statement in the L\"{u}ders theorem.

There are strong indications that the L\"{u}ders theorem extends to the case
of unsharp observables: in fact two important special cases have been proven
recently \cite{BS98}. It can be argued that these two cases are sufficiently
comprehensive for physical purposes. They will serve the present needs. The
formulation of the following proposition rests on the notion of a L\"{u}ders
measurement for unsharp observables. Let $A$ be an unsharp observable
represented by a complete family of effects, $A=\{E_{i}\}_{i=1...N}$, 
$\sum_iE_i=I$. A nonselective L\"uders measurement of $A$ leads to a state
change of the object that is described by the L\"{u}ders state
transformation, defined via 
\[
\rho \longmapsto \mathcal{I}_{L}^{A}(\rho ):=\sum_{i=1}^{N}E_{i}^{1/2}\rho
E_{i}^{1/2}
\]
for all state operators $\rho $ (for details on this concept, cf.\ \cite
{BGL95}). One considers the question under what conditions the outcome of a
measurement of an effect $B$ does not depend on whether or not a
nonselective measurement of $A$ has been carried out.

\begin{proposition}
\label{p1}Let $A=\{E_{i}\}_{i=1...N}$ be a collection of effects such that 
$\sum_{i}E_{i}=I$, and let $B$ be an effect. Then the equivalence 
\[
\lbrack B,E_{i}]=0\quad \forall i\qquad \Longleftrightarrow \qquad \mathrm{tr}
[\rho B]=\mathrm{tr}[\mathcal{I}_{L}^{A}(\rho )B]\quad \forall \rho
\]
holds (at least) in the following two cases:\newline
\textrm{(}$\alpha $\textrm{)} $B$ has a discrete spectrum of eigenvalues
that can be ordered in decreasing order, $A$ arbitrary;\newline
\textrm{(}$\beta $\textrm{) }$A=\{E_{1,}I-E_{1}\}$, $B$ arbitrary.
\end{proposition}

In order to formulate weak Einstein causality, we need to assume a
physically meaningful association of observables with (bounded open)
spacetime regions, in the sense that such observables can be measured by
means of operations  carried out within these regions. Such
measurements, operations and observables will be called \textsl{local}. 
In the context of algebraic relativistic quantum  theory, a measurement is local
if the operations representing the associated state changes are expressible
in terms of elements of the corresponding local algebra of observables \cite{dMu}. This
condition is satisfied for L\"uders measurements.
Now,  
\textsl{weak Einstein causality} means the following: if $A$ and $B$
represent observables associated with two spacelike separated spacetime 
regions $R_{1}$, $R_{2}$,
respectively, then the act of a nonselective local measurement of $A$
should not influence the outcomes of a measurement of $B$ (and vice versa).
Hence, weak Einstein causality
for L\"uders measurements of local observables reads: 
\begin{equation}
R_{1},R_{2}\text{\ spacelike separated\quad }\Longrightarrow\quad\mathrm{tr}[\rho
B]=\mathrm{tr}[\mathcal{I}_{L}^{A}(\rho )B]\;\forall \rho .  \tag{C}
\end{equation}
 \noindent The state description $\mathcal{I}_{L}^{A}(\rho )$ is
appropriate to local observers acting in $R_2$ since -- given that
there are no classical signals faster than light -- it represents
the information available to them if they know that a measurement in
$R_1$ is taking place. 
In view of Proposition 1, it follows that for local
measurements of unsharp observables in spacelike separated regions to
satisfy weak Einstein causality, these observables must commute. This is
remarkable as local measurements at spacelike separations can be
regarded in a way as joint measurements, and it is known that unsharp
observables can be jointly measurable without necessarily commuting with each other.

Schlieder's argument can now be formulated for localization observables as
follows.

\begin{proposition} \label{p2} Let $A=\{E_{1},E_{2}\}$, where
$E_{1}=E_{\Delta _{1}},\ E_{2}=I-E_{\Delta _{1}}$ and $B=E_{\Delta _{2}}$
are localization effects, and $\Delta _{1}\in
\mathcal{F}(S_{1})$, $\Delta _{2}\in\mathcal{F}(S_{2})$ are bounded spatial
sets contained in spacelike separated regions $R_{1},R_{2}$,
respectively. Suppose that $A$ and $B$ are locally measurable in these
regions and that the Einstein causality condition (C) holds for them. 
Then $E_{\Delta_{1}}$ and $E_{\Delta_{2}}$ commute.
 \end{proposition}
This statement is an immediate consequence of case ($\beta $) of Proposition 
\ref{p1}. It is interesting to observe the following consequence: if a
localization observable $\Delta \mapsto E_{\Delta }$ is represented as a 
\textsc{pom} and measurable by means of local operations, 
then causality requires this \textsc{pom} to
be commutative since all bounded disjoint spatial sets of the 
$\sigma $-algebra $\mathcal{F}(S)$ are spacelike separated and hence commute with
each other. On the other hand, covariant relativistic phase space
observables are constructed via generalized coherent states and thus are
noncommutative (see \cite{Ali85,Ali98}). Hence, such phase space
observables cannot satisfy local commutativity.
This is not too surprising as a phase space localization
measurement involves a measurement of momentum which itself is not a local
observable. However, there is a possibility that the spatial marginals of
a suitable covariant family of phase space observables are commutative.

Besides weak causality, there is another, independent requirement that
entails local commutativity. This is the condition of (relativistic)
\textsl{objectivity}: the descriptions of
a pair of spacelike separated local measurements given by different
inertial observers should be consistent with each other. Thus, consider two
inertial observers in the intersection of the forward lightcones (causal
influence regions) of the two spacelike separated regions $R_1,R_2$. 
Suppose the observers are moving relative to each other in such a way
that they assign different time orderings to the two local measurements.
Relativistic objectivity means that they predict and record the same statistics for
all possible future measurements.
Therefore, although the time orderings are different,
the successive state changes should nevertheless lead to the same final
state in both descriptions. Let $A=\{E_1,E_2,\dots\}$, $\sum E_i=I$, and
$B=\{F_1,F_2,\dots\}$, $\sum F_j=I$ represent the two discrete local
observables in question, and let $\mathcal{I}^A_{L,i}$,
$\mathcal{I}^B_{L,j}$ denote the associated L\"uders operations. Then objectivity
is expressed by the following condition \cite{S68}:
$$
\aligned
R_{1},R_{2}\text{\ spacelike separated\quad}\Longrightarrow\quad
&\mathcal{I}_{L,i}^{A}\left( \mathcal{I}_{L,j}^{B}(\rho )\right) =
\mathcal{I}_{L,j}^{B}\left( \mathcal{I}_{L,i}^{A}(\rho )\right) \cr
&\mathrm{for\ all\ } \rho,i,j. \endaligned 
\eqno({\rm O}) 
$$
It can be shown that the commutativity of the L\"uders operations for
$A$ and $B$ is equivalent to the commutativity of all $E_i$ with all
$F_j$ \cite{Bu99}.
(For pairs of observables in the domain of applicability of Proposition
1, this equivalence is a consequence of the fact that due to
 $\sum_i\mathcal{I}_{L,i}^{A}=\mathcal{I}_{L}^{A} $, the commutativity
property in (O) implies the nondisturbance property in (C)). Therefore,
the relativistic objectivity condition is equivalent, via local
commutativity, to weak Einstein causality. This important
connection (e.g., \cite{Mit98} and references therein) is thus found to
be valid also for unsharp observables. The fact that the class of
measurements used in these arguments is of the  L\"{u}ders type is only
of technical significance: weak causality and objectivity would be 
necessary requirements to be imposed on \textsl{all} local measurements and
therefore, in particular, on L\"{u}ders measurements.

The potential conflict between causality and localizability for
relativistic quantum mechanics has been highlighted by Hegerfeldt 
on the basis of a notion of strong causality that requires subluminal
propagation of localization properties (for a concise review, see \cite{H98}). 
More precisely, \textsl{strong causality }requires that for a bounded
spatial set $\Delta $, if the probability of localization at time $t=0$
is 1, then the probability of localization at time $t>0$ within the
inflated set $\Delta _{t}= \{x\in  S\,|\,\mathrm{dist}(x,\Delta )\leq
ct\}$ is also equal to 1. It is shown that this condition cannot be
satisfied by any (covariant) localization observable. 

It is not hard to see that in the case of sharp localization observables
represented as a normalized projection valued measure $\Delta \mapsto
P_{\Delta }$, the strong causality implies local commutativity (3) and thus weak
causality.  Translating this into the Heisenberg picture, this means
$P_{\Delta }\leq P_{\Delta _{t}+ta}$, where $a$ is the future
timelike unit vector perpendicular to $S$. Now let $\Delta _{1}\in S$,
$\Delta _{2}\in $ $S+ta$ be two spacelike separated bounded spatial
sets. Strong causality yields 
$ P_{\Delta _{1}}\leq P_{\Delta_{1,t}+ta}\leq I-P_{\Delta _{2}}$, 
and so $P_{\Delta _{1}}P_{\Delta_{2}}=0$. 
Thus, the violation of local commutativity for sharp localization
observables entails the violation of strong causality. It should be
noted, however, that Hegerfeldt's theorem takes into account the
possibility of localization operators $E_{\Delta }$ which are effects
but not projections. In this case strong causality implies the following
chain of inclusions for the respective spectral projections:  
\[
P_{\Delta _{1}}^{(1)}\leq P_{\Delta _{1,t}+ta}^{(1)}\leq P_{\Delta
_{2}}^{(0)}\leq I-P_{\Delta _{2}}^{(1)}, 
\]
 and so $P_{\Delta_{1}}^{(1)}P_{\Delta _{2}}^{(1)}=0$. 
It is not clear whether in this
case strong causality does imply weak causality (for localization effects).
But the assumption of local commutativity, via Theorem \ref{t3}, is seen to render
strong causality inapplicable: the premise of strict initial localization
cannot be fulfilled. Therefore, in
the present framework strong causality is an unnecessary assumption. Its
virtue lies in the fact that some of the other postulates (such as
covariance and even the group representation) could be dropped and still
strong causality is violated in the sense of instantaneous
delocalization of wave functions \cite{H98}. However, without 
translation covariance -- which we regard as a defining property of localization
observables -- it appears doubtful whether such delocalizations can be
interpreted as giving rise to (possible) superluminal particle propagations.

\section{Discussion}

We have generalized a theorem due to Schlieder which states the
incompatibility between local commutativity and  sharp covariant
localizability. Based on a generalization of L\"{u}ders' theorem, we
then found that local commutativity is equivalent to weak Einstein
causality as well as to a postulate of relativistic objectivity, also in
the case of unsharp observables -- provided they admit local measurements.
The results of Sections 2 and 3 leave us with the following situation.

 (I.) For localization observables $\Delta\mapsto E_{\Delta}$  admitting
sharply localized states (i.e., states which yield probability 1 in
bounded spatial  sets), local commutativity is violated (Theorems 1 and
3). If there is any physical sense in saying that the effects $E_{\Delta}$ are
measurable by local operations, then weak Einstein causality would be
violated (Propositions 1 and 2).  Such local, sharp localization
measurements would lead to  statistical influences between spacelike
separated regions and hence, superluminal signals. 

(II.) For localization observables to satisfy local commutativity, they
must necessarily be strongly unsharp. It is an open question whether 
local commutativity is actually satisfied for any or all strongly unsharp
localization observable; an interesting (though unlikely) class of candidates to be
investigated is given by the spatial marginals of phase space
observables. If a strongly unsharp localization observable can be
measured by local operations, and if it violates local commutativity,
this would again imply a violation of weak Einstein causality. 

The problematic conclusion in (I.) can be countered by arguing that
sharp spatial localization is an operationally meaningless idealization.
It would in fact seem implausible to ignore the quantum nature of the
constituents of the detectors used to define spatial localization sets.
It appears more likely that realistic procedures for measuring
localization observables, which are based on quantum probes with
extended wave functions and interactions with infinite ranges (albeit
with decreasing strengths), would render the spatial localization sets
intrinsically fuzzy. Also, confining a quantum object or probe within
sharp spatial boundaries would require an infinite amount of energy,
(e.g., an infinite potential well). Thus, the causality violation for
sharp localizations could be seen as an artefact arising from an
unjustified idealization. (From the perspective of a relativistic
quantum field theory, such strong interactions as would be needed for
sharp localization measurements would  give rise to particle pair
creations and could not therefore be described within a single-particle
theory. However, we would prefer a decision on the limitations of a
single-particle relativistic quantum mechanics from \textsl{within} that
theory.)

The conclusion of causality violations for either sharp localizations
(scenario (I.)) or strongly unsharp localizations violating local
commutativity (scenario (II.)) may be barred for another, common reason:
localization observables could simply fail to be local observables. In
fact, the concept of a localization observable -- whether sharp or
unsharp -- involves global elements, namely, the totality of all bounded
spatial subsets of $S$ as well as the defining requirement of
translation covariance. It is thus quite conceivable that from their
very operational definition, sharp or unsharp localization effects
cannot be regarded as locally measurable quantities. As local
measurability is a premise of the weak Einstein causality postulate (and
of the objectivity requirement), this postulate and, along with it,
Proposition 2 would in this situation become  
inapplicable. Thereby the validity of weak
causality would not be affected by a violation of the
commutativity condition (3); indeed this condition
would lose its intended meaning indicated by the phrase `local
commutativity'. This interpretation of Theorems 1 and 3 is in accordance
with the analysis given by Schlieder \cite{S71}: in the context of a
relativistic local quantum (field) theory, which stipulates local
commutativity, Theorem 1 is interpreted as implying that localization 
observables admitting sharp localization cannot belong to a local
algebra. 

The assumption of local measurability of localization observables has
also been challenged by Butterfield and Fleming \cite{BF98} in a recent
lucid analysis of the `strange' properties of localizations. I feel this
point requires a detailed measurement theoretic investigation before
Schlieder's theorem could be taken as conclusive evidence against the
possibility of a relativistic quantum mechanics of particles in the
sense proposed by Malament \cite{Mal96}.

If it is granted that localization observables allowing sharply
localized states are not local observables, so that weak causality would
not be challenged, then Hegerfeldt's theorem would still entail the
instantaneous spreading of wave functions and hence potential
superluminal propagation of a particle. I would argue that this cannot be
used for superluminal signalling: In order for the particle to carry a
bit of information, the sender would need to be able to control the
particle to the extent that (s)he is capable of either releasing the
particle or keeping it trapped. This would entail the infinite energy
problem mentioned above. On the other hand, if the particle is free, the
`sender' will have no control over it, and the fact that it had been
localized in a bounded region at some initial instant of time does not 
by itself carry any information to a spacelike separated receiver.

The preceding lines of reasoning can be carried somewhat further. Up to
now we have assumed that bounded spacetime regions -- in which local
physical operations are to be carried out -- can be operationally
defined solely by classical physical means. Thus it is assumed that the
quantum nature of the constituents of the relevant measuring devices can
be ignored. It would be the task of a relativistic quantum theory of
spacetime measurements, which remains yet to be developed, to justify
this assumption. As indicated above, sharp localization may not be an
operationally meaningful concept; given that the (quantum) devices used 
to define spacetime regions are themselves only unsharply localizable, 
the concept of a local measurement -- and with it that of a local
observable algebra -- would have to be reformulated. This would render
weak and strong Einstein locality inapplicable and would call for an
operationally significant notion of causality,  possibly in
the form of a probabilistic concept and involving reference to 
appropriate levels of detector sensitivities.

Hegerfeldt has proposed a probabilistic causality condition  according
to which the `bulk' of a wave packet propagates with  subluminal speed.
This  condition is still found to be violated for approximately
localized states with fast decaying (`exponential') tails \cite{H98}. On
the other hand, it has been shown that the probabilities for causality
violating behavior are spurious from a practical point of view (e.g.,
\cite{GP84}). However, the experimental relevance of the present
formulations of probabilistic causality does not seem entirely evident;
in particular, it is again unclear whether violations of this 
causality condition would permit the existence of superluminal signals.

Finally we point out that a coherent account of the status of postulates
such as local commutativity, or weak and strong causality in current
relativistic quantum theories is needed not only for the sake of
theoretical argument. In recent years, various experimental groups have
reported demonstrations of superluminal propagation phenomena with
evanescent microwave modes and with pairs of photons passing through
opaque media. An up to date discussion of such experiments can be found
in the special Proceedings issue of Annalen der Physics cited in Refs.~
\cite{H98,Mit98}. Some authors seem close to suggesting that the
possibility of superluminal signalling on the basis of such experiments
may not be ruled out. Such a claim  cannot be refuted simply by making
reference to the fact that our existing relativistic quantum theories
incorporate the principle of weak causality: they do so because it has
been built in `by hand', namely by stipulation of local commutativity.
This situation raises the interesting question as to what could
constitute a principal theoretical demonstration of the impossibility of
superluminal signals. (It appears to me that it would be hard to dismiss
relativistic objectivity, which seems to constitute a necessary
precondition for any scientific theory; and as we have seen, within
Hilbert space quantum theory this implies local commutativity and weak
causality.)  As long as a plain experimental demonstration of
superluminal signals is lacking (which I suspect it will be for a very
long time), the compatibility of these experiments with relativistic
causality could only be demonstrated by way of providing a satisfactory
quantitative account of the experiments using relativistic quantum
theory; this would amount to giving a causal explanation. Such a study
will have to be based on the use of sound localization observables for
the photons involved; and these are known to be necessarily strongly
unsharp \cite{BrS96}. Thus, all the questions raised above concerning the
(non-)locality and possible probabilistic causality of localizations
will have to be addressed. Some of these issues  will be taken up in a
forthcoming study by J.\ Brooke and the author \cite{BB98}.

\section*{Acknowledgement} It is a pleasure to thank James Brooke for
helpful comments on a previous version of the manuscript.


\section*{\protect\vspace{1pt}Appendix 1: Effects, POM's, Unsharpness,
and All That}

\emph{Operational Quantum Theory }is a conceptual completion of quantum
mechanics on Hilbert space ($\mathcal{H}$) in the sense that the most
general notion of observable is incorporated that is compatible with the
probabilistic structure of the theory. An observable encompasses the
totality of statistics of a given measurement (or class of measurements
yielding the same statistics) with respect to all input states. Insofar as
the probability for an outcome is to be given in terms of the expectation
value of some operator in any state, it follows that such an operator must
be positive and have a spectrum within the interval $\lbrack 0,1\rbrack$.
Operators of this kind, which represent the occurrence of a particular
outcome in a measurement, are called \emph{effects}.
Hence, a linear operator $E$ is an effect if it is bounded by $O$
and $I$, $O\le E\le I$.  An observable will
thus be constituted by an association of effects with subsets of possible
values. The probability for an outcome in one of a collection of disjoint
value sets should be given by the sum of probabilities for the separate
sets. This leads to the additivity property of the map from value sets to
effects. For convenience, one allows additivity for finite or countably many
disjoint sets and refers to this as $\sigma $-additivity. In addition, it is
usually assumed that for a given measurement it is certain that \emph{some}
outcome will occur, that is, that the probability for the total value set, 
$\Omega $,  is equal to unity. This is to say that the associated effect in
the unit operator, $I$. One is thus led to the definition of an observable
as a normalized positive operator valued (or effect valued) measure $\Sigma
\ni X$\noindent $\longmapsto E(X)$ defined on a measurable space $\left(
\Omega ,\Sigma \right) $ such that:
\begin{eqnarray*}
O &\leq &E(X)\quad \ \ \ \ \ \ \ \ \ \ \ \ \ \ \ \ \ \ \ \ \ \ \ \ \ \ \ \ \
\ \ \ \ \ \ \ \ \ \ \ \ \ \ \ \ \ \text{(positivity)} \\
E(\bigcup_{i}X_{i}) &=&\sum_{i}E(X_{i})\quad \text{if }X_{j}\cap
X_{k}=\emptyset ,\text{ }j\neq k\text{\ \ \ \ \ \ \  \ \ \ \ (}\sigma \text{-additivity)}
\\
E(\Omega ) &=&I\text{ \ \ \ \ \ \ \ \ \ \ \ \ \ \ \ \ \ \ \ \ \ \ \ \ \ \ \
\ \ \ \ \ \ \ \ \ \ \ \ \ \ \ \ \ \ \ \ \ \ \ \ \ \ (normalization) }
\end{eqnarray*}
It should be noted that observables in the usual sense are captured by this
definition by virtue of the spectral theorem: first of all, the set of
effects contains all projections; and any self-adjoint operator gives rise
to a unique projection valued measure on the real Borel algebra $\left( 
\mathcal{B}({\Bbb R})\right) $, its spectral  measure. Effect valued measures that
are not projection valued are distinguished from projection valued measures
by the following important property: a projection $P$ and its complement, 
$P^{\perp }:=I-P$, are  necessarily orthogonal to each other, that is, 
$PP^{\perp }=O.$ Conversely, if an effect $E$ and its complement $E^{\perp
}:=I-E$ satisfy $EE^{\perp }=O$ then $E$ is a projection (as $E=E^{2}$).
This fact suggests the definition of a \emph{sharp observable} as an effect
valued measure such that for any effect $E$ in its range, $E$ and its
complement $E^{\perp }$ have no common positive, nonzero lower bound; this
is equivalent to $EE^{\perp }=O$, and hence to the statement that the effect
valued measure is actually a projection valued measure. All other
observables will be referred to as \emph{unsharp observables}. Note that it
is possible for an unsharp observable  to have definite values, namely, if
the effect associated with some value or range of values has eigenvalue 1.
The term \emph{unsharpness} only refers to the fact that some effects in the
range of the observable have a spectrum within $\lbrack 0,1\rbrack $ not
limited to (a subset of) $\{0,1\}$. For further information on unsharp
observables, the reader may refer to the monograph \cite{BGL95}.

\section*{Appendix 2: Proof of Theorem \ref{t3}}

We follow Malament (1996)\strut\ step by step, the only modification being
an appropriate choice of the pairs of projections in question. The reader
may wish to accompany the construction with a spacetime diagram or else
consult Figure 1 of (Malament, 1996). The covariance properties of our
projections are a consequence of the corresponding covariance properties of
the effects and the spectral theorem. Let $\Delta \in \mathcal{F}(S)$ be a
bounded spatial set and $a\in M$ tangent to $S$ such that $\Delta $ and 
$\Delta +a$ are disjoint and such that for all future directed timelike $a_{1}
$ and all sufficiently small (in modulus) $t$, $\Delta $ and $\Delta
+a+ta_{1}$ are spacelike separated. Due to the localization condition (2')
one has 
\[
\forall \varphi \in \mathcal{H}\quad (\Vert \varphi \Vert =1):\quad \langle
\varphi |E_{\Delta }\varphi \rangle =1\Longrightarrow \langle \varphi
|E_{\Delta +a}\varphi \rangle =0.
\]
This is equivalent to 
\begin{equation}
P_{\Delta }^{(1)}(I-P_{\Delta +a}^{(0)})=0.  \tag{i}
\end{equation}
Applying \ translation covariance and locality, one obtains for sufficiently
small $t$: 
\[
\lbrack E_{\Delta },U(ta_{1})E_{\Delta +a}U(ta_{1})^{\ast }\rbrack =\lbrack
E_{\Delta },E_{\Delta +a+ta_{1}}\rbrack =0,
\]
and thus also (by virtue of the spectral theorem) 
\begin{equation}
\lbrack P_{\Delta }^{(1)},U(ta_{1})(I-P_{\Delta +a}^{(0)})U(ta_{1})^{\ast
}\rbrack =\lbrack P_{\Delta }^{(1)},I-P_{\Delta +a+ta_{1}}^{(0)}\rbrack =0. 
\tag{ii}
\end{equation}
Now apply Lemma \ref{l1} \lbrack taking $V(t)=U(ta_{1})$, $E_{1}=E_{\Delta }$,
$E_{2}=I-P_{\Delta +a}^{(0)}$\rbrack\ -- using the spectrum condition and
(i), (ii) -- to conclude that for all future directed timelike $a_{1}$, and
all $t$, $P_{\Delta }^{(1)}U(ta_{1})(I-P_{\Delta +a}^{(0)})U(ta_{1})^{\ast
}=0$, and therefore 
\begin{equation}
P_{\Delta }^{(1)}(I-P_{\Delta +a+ta_{1}}^{(0)})=0.  \tag{iii}
\end{equation}
Next let $a_{2}$ be any future directed timelike unit vector. For
sufficiently large $t_{2}>0$, the set $\Delta +t_{2}a_{2}$ is to the
timelike future of $\Delta +a$. Then one can find $t_{2}>0$ and $\varepsilon
>0$ such that $\Delta +(t_{2}+t)a_{2}$ is to the timelike future of $\Delta
+a$ for all $t$ with $|t|<\varepsilon $. Hence if $|t|<\varepsilon $, there
is a future directed timelike unit vector $a_{1}$ and a number $t_{1}$ such
that $\Delta +(t_{2}+t)a_{2}=\Delta +a+t_{1}a_{1}$. Thus, by (iii), if 
$|t|<\varepsilon $, then $P_{\Delta }^{(1)}(I-P_{\Delta
+(t_{2}+t)a_{2}}^{(0)})=0$, or equivalently, by translation covariance: 
$P_{\Delta }^{(1)}U(ta_{2})(I-P_{\Delta +t_{2}a_{2}}^{(0)})U(ta_{2})^{\ast }=0
$. Invoking Lemma \ref{l1} again \lbrack with $E_{1}=$ $P_{\Delta }^{(1)}$, 
$E_{2}=I-P_{\Delta +t_{2}a_{2}}^{(0)}$\rbrack , one obtains 
\[
P_{\Delta }^{(1)}U(ta_{2})(I-P_{\Delta +t_{2}a_{2}}^{(0)})U(ta_{2})^{\ast }=0
\]
and hence (by translation covariance) 
\[
P_{\Delta }^{(1)}U\left( (t+t_{2})a_{2}\right) (I-P_{\Delta }^{(0)})U\left(
(t+t_{2})a_{2}\right) ^{\ast }=0
\]
for \textsl{all} $t$. Choosing $t=-t_{2}$, one concludes that $P_{\Delta
}^{(1)}(I-P_{\Delta }^{(0)})=0$, and so $P_{\Delta }^{(1)}\leq P_{\Delta
}^{(0)}$, that is, 
\[
P_{\Delta }^{(1)}=0.
\]


\end{document}